\documentclass[letter,useAMS,usenatbib]{mn2e}

\usepackage{amsmath}
\usepackage{amssymb}
\usepackage{graphicx} 
\usepackage{sansmath}
\usepackage{flafter}
\usepackage{float}
\usepackage{fixltx2e}
\usepackage{url}
\usepackage{hyperref}

\newcommand{\Omb}{\Omega_\mathrm{b}}

\newcommand{\ROLR}{R_{\mathrm{OLR}}}

\newcommand{\vOLR}{v_{\mathrm{OLR}}}

\newcommand{\Kpc}{~\mathrm{kpc}}

\newcommand{\kmsec}{~\mathrm{km}~\mathrm{s}^{-1}}

\newcommand{\kmseckpc}{~\mathrm{km}~\mathrm{s}^{-1}~\mathrm{kpc}^{-1}}

\newcommand{\vc}{v_{\mathrm{c}}}

\newcommand{\RNum}[1]{\uppercase\expandafter{\romannumeral #1\relax}}

\newcommand{\degree}{^{\circ}}

\newcommand{\vlos}{v_\mathrm{los}}

\newcommand{\mas}{\mathrm{mas}}

\newcommand{\pare}[1]{\left(#1\right)}
\newcommand{\paresq}[1]{\left[#1\right]}

\newcommand{\Eq}[1]{Eq.~(\ref{#1})}

\newcommand{\Fig}[1]{Fig.~\ref{#1}}

\newcommand{\phib}{\phi_{\mathrm{b}}}

\title[New evidence for a fast bar in the Milky Way]{Tracing the \emph{Hercules} stream with Gaia and LAMOST: new evidence for a fast bar in the Milky Way}

\author[G.~Monari~et~al.]{
Giacomo Monari,$^{1,2}$\thanks{E-mail: giacomo.monari@fysik.su.se}
Daisuke Kawata,$^{3}$
Jason A. S. Hunt$^{3,4}$
and Benoit Famaey$^{1}$
\\
$^1$ Universit\'e de Strasbourg, CNRS, Observatoire astronomique de Strasbourg, UMR 7550, F-67000 Strasbourg, France\\
$^2$ The Oskar Klein Centre for Cosmoparticle Physics, Department of Physics, Stockholm University, AlbaNova, 10691 Stockholm, Sweden \\
$^{3}$ Mullard Space Science Laboratory, University College London, Holmbury St. Mary, Dorking, Surrey, RH5 6NT, UK\\
$^{4}$ Dunlap institute for Astronomy and Astrophysics, University of Toronto, 50 St. George Street, Toronto, Ontario, M5S 3H4, Canada
}

\date{Accepted XXX. Received YYY; in original form ZZZ}

\pubyear{2016}

\begin{document}
\label{firstpage}
\pagerange{\pageref{firstpage}--\pageref{lastpage}}
\maketitle

\begin{abstract}
The length and pattern speed of the Milky Way bar are still controversial. Photometric and spectroscopic surveys of the inner Galaxy, as well as gas kinematics, favour a long and slowly rotating bar, with corotation around a Galactocentric radius of 6~kpc. On the other hand, the existence of the \emph{Hercules} stream in local velocity space favours a short and fast bar with corotation around 4~kpc. This follows from the fact that the \emph{Hercules} stream looks like a typical signature of the outer Lindblad resonance of the bar. As we showed recently, reconciling this local stream with a slow bar would need to find a yet unknown alternative explanation, based for instance on the effect of spiral arms. Here, by combining the TGAS catalogue of the Gaia DR1 with LAMOST radial velocities, we show that the position of \emph{Hercules} in velocity space as a function of radius in the outer Galaxy indeed varies exactly as predicted by fast bar models with a pattern speed no less than 1.8 times the circular frequency at the Sun's position. 
\end{abstract}




\section{Introduction}

The distribution and kinematics of stars in the Milky Way disc can be
described, to a first approximation, by axisymmetric models \citep[e.g.][]{BinneyPiffl2015}.
However, it is well known that the Galactic disc contains
non-axisymmetric features. These appear in the distribution of stars
in configuration space: for example, the Milky Way disc contains a
prominent bar \citep[e.g.][]{Binney1997} and a still relatively poorly known 
spiral arm structure in both gas and stars \citep[e.g.][]{Benjamin2005,Vallee2008}. However, non-axisymmetries also reveal themselves in the velocity
distribution of stars, which can help learning about their dynamical properties 
such as their pattern speed. In particular, the velocity distribution of stars in the
Solar neighbourhood would appear as an homogeneous ellipsoid, with a
tail for low tangential velocities, if the Galaxy were a pure axisymmetric disc
in differential rotation. However, a number of substructures in local velocity space
space have been known for a very long time \citep{Proctor1869,Kapteyn1905,Eggen1958}. The astrometric and spectroscopic surveys of the late XXth and early XXIst century \citep{Hipparcos,GCS,Famaey2005} have clearly confirmed that the kinematics of disc stars in the Solar neighbourhood
is rather clumpy, and that the most prominent of these clumps (known as `moving groups') are not disrupted
open clusters keeping coherence in velocity
space. An alternative mechanism to create moving groups (or dynamical streams) 
in local velocity space is the resonant interaction between the stars and the main non-axisymmetric patterns of the Milky Way, namely the bar and the spiral arms. In particular,
\citet{Dehnen1999,Dehnen2000} pointed out that the \emph{Hercules} moving
group, or \emph{Hercules} stream, could be a direct consequence of the Sun being located just outside of the bar's outer Lindblad resonance (OLR). This resonance occurs at the
radius $\ROLR$ where stars make two epicyclic oscillations while making one retrograde rotation in the frame of the bar, hence
\begin{equation}\label{eq:OLR}
  \kappa+2\pare{\Omega-\Omb}=0,
\end{equation}
where $\Omega(R)$, $\kappa(R)$, and $\Omb$ are the Galaxy's circular
frequency, epicyclic frequency, and the bar's pattern speed
respectively \citep[see][]{BT2008}.

However, careful studies of the photometry of the inner Galaxy \citep{WeggGerhard2013, Wegg2015} 
recently revealed a long and thin extension of the bar, oriented at an angle of
$\phib \sim 27\degree$ from the Galactic centre-Sun direction, and
reaching a Galactocentric radius $R \sim 5$~kpc. If this structure is not a loosely wound spiral arm connected to the bar but an extension of the bar itself,
this limits its pattern speed since it cannot extend beyond corotation. 
Such a long and slow bar has subsequently been vindicated by recent dynamical modelling
of stellar and gas kinematics in the inner Galaxy \citep{Portail2015,Sormani2015,Li2016,Portail2016},
converging to a pattern speed of the order of $40 \kmseckpc$
placing the bar corotation at about 6~kpc from the Galactic centre \citep{Portail2016}, and the OLR way beyond the Solar neighbourhood.

This is clearly at odds with the structure of local velocity space. Indeed, as we showed recently \citep{Monari2016c}, a slow bar with a pattern speed as above does not produce a bimodality in local velocity space. Hence, to reconcile this feature with a slow bar, an alternative explanation, based, e.g., on spiral arms, should be found. For instance, \citet{Grand2014} showed that the outward radial migrators behind their corotating spiral arms display lower tangential velocities and an outward velocity. 

As pointed out by, e.g. \citet{Bovy2010}, a way to check whether the OLR explanation holds is to investigate how the \emph{Hercules} feature in velocity space varies with the position in the Galaxy. This was already shown to hold well in the relatively nearby inner Galactic disc by \cite{Antoja2014}, using the RAVE survey \citep{Steinmetz2006}. However, the advent of Gaia \citep{Gaia} provides a unique opportunity to test this in the outer Galactic disc. In particular, cross-matching the recent TGAS catalogue with existing spectroscopic surveys could allow us to trace the position of \emph{Hercules} in velocity space as a function of Galactocentric radius. In this {\it Letter}, we show that this can be achieved through the combination of TGAS with the LAMOST DR2 catalogue \citep{Liu2014}.

\section{Cross matching TGAS and LAMOST}

We cross match the TGAS catalogue (from which we obtain the Right Ascension $\alpha$, the declination $\delta$, the parallax $\pi$, and the proper motions $\mu_\alpha$ and $\mu_\delta$) with the LAMOST DR2 A, F, G and K type stars catalogue \citep[][from which we obtain the line--of--sight velocity $\vlos$]{Liu2014} using Gaia tools written by Jo Bovy\footnote{\url{http://github.com/jobovy/gaia_tools}.}. The LAMOST A, F, G and K type stars catalogue selects only the spectra whose signal-to-noise ratio (SNR) in the $g$-band is greater than 6 in dark time and 20 in bright time. We obtain in this way 108,910 stars. We then select stars with fractional parallax error\footnote{We take into account the recommended value for the systematic error in the parallax of $0.3$ $\mas$ \citep[see][]{Gaia}.} $\sqrt{\sigma_{\rm \pi}^2+(0.3 \mas)^2}/\pi<0.2$, which provides us a total 49,075 stars, mostly distributed towards the anti--centre of our Galaxy (while, RAVE, for example, focuses on the central parts). These parallax errors ensure a very limited bias on distances when inverting them. In the following analysis, we used all the stars satisfying the above criteria to maximise our sample. The cross-matched TGAS and LAMOST dataset has a complicated selection function. However, because we focus on the velocity field of relatively nearby stars (distances $\lesssim 0.7~\Kpc$), it is unlikely that our conclusions are affected by the observational selection function.

\section{Analysis}

\begin{figure*}
  \centering 
  \includegraphics[width=0.65\columnwidth]{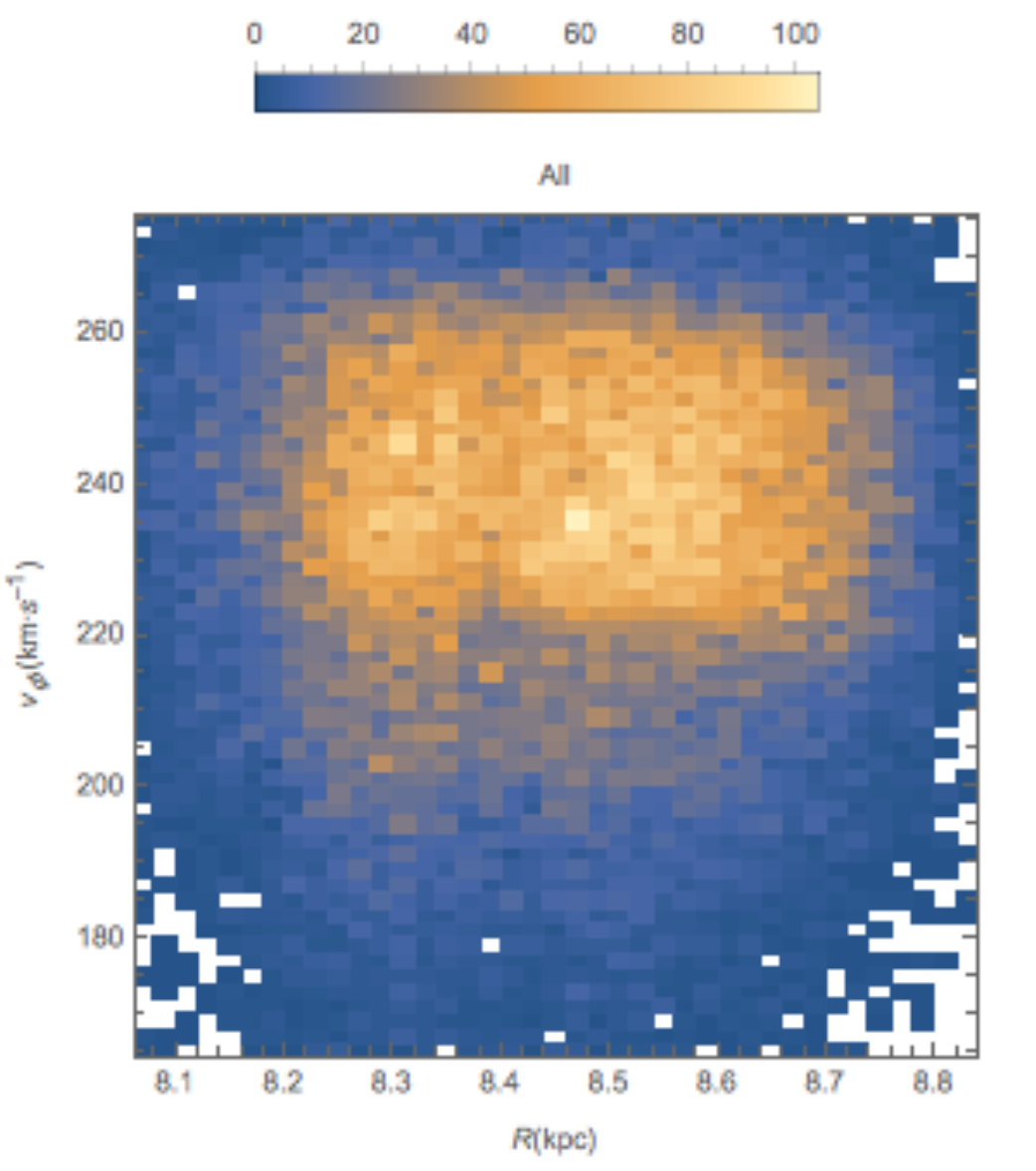}
  \includegraphics[width=0.65\columnwidth]{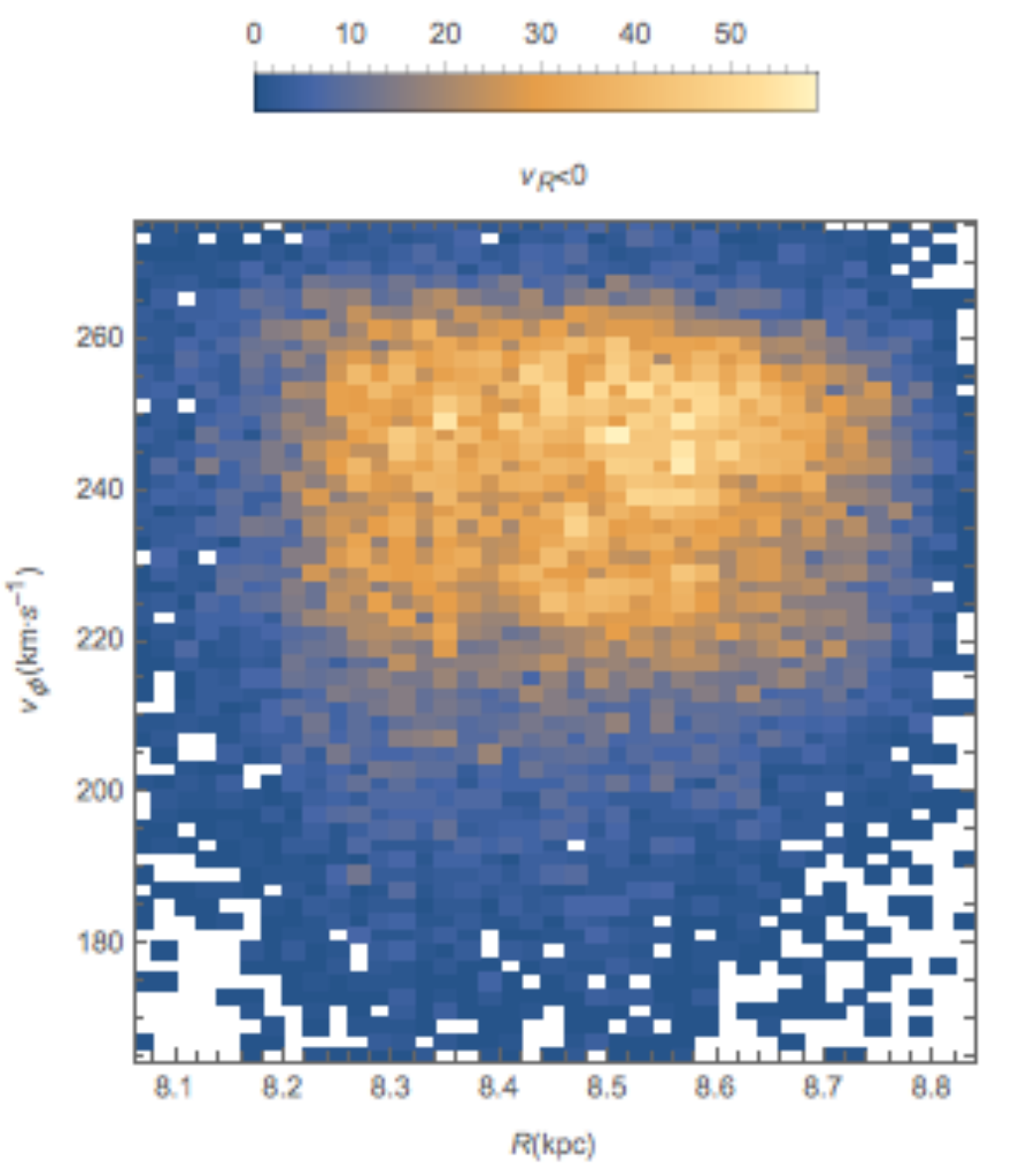}
  \includegraphics[width=0.65\columnwidth]{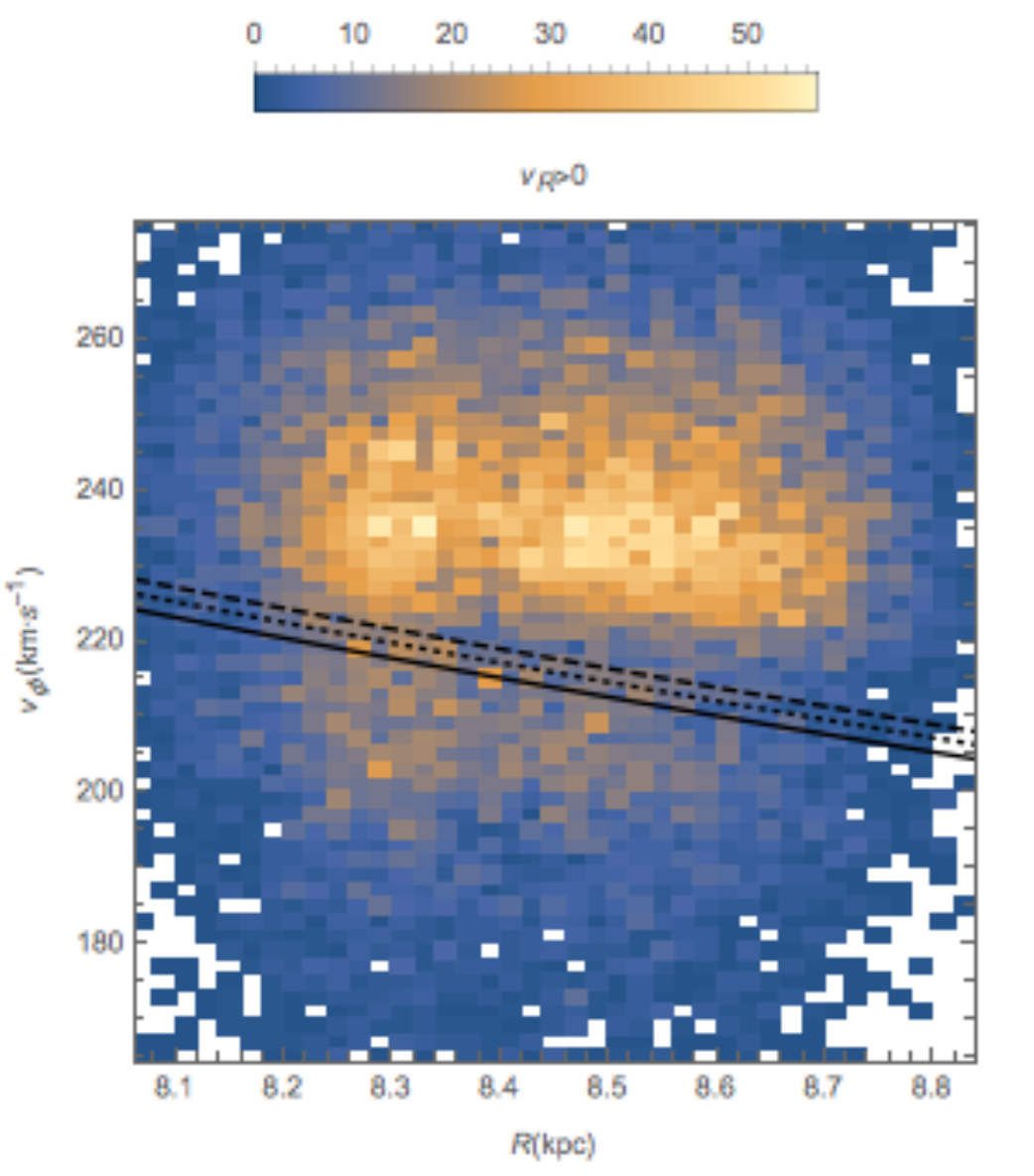}
  \caption{Distribution of stars in TGAS+LAMOST in the $(R,v_\phi)$ plane. We select stars with parallax accuracy $\sqrt{\sigma_\pi^2+(0.3~\mas)^2}/\pi<0.2$. The bin size is 20~pc in $R$, and $2~\kmsec$ in $v_\phi$, and the units of the color bar indicate the number of stars per bin (the white bins are empty). The left panel represents the whole sample, the central panel stars with $v_R<0$, and the right panel stars with $v_R>0$. The different curves in the right panel correspond to different models of $\vOLR$ with $\Omb=1.89\Omega_0$: the solid curve has a flat $\vc(R)$ ($\beta=0$), the dashed curve has an increasing $\vc(R)$ ($\beta=0.3$), and the dotted curve a decreasing $\vc(R)$ ($\beta=-0.3$). Note how \emph{Hercules} fades away beyond $R=8.6\Kpc$, in accordance with \Fig{fig:mod}.}
\label{fig:Rvphi_cut}
\end{figure*}
\begin{figure*}
  \centering 
 \includegraphics[width=0.8\textwidth]{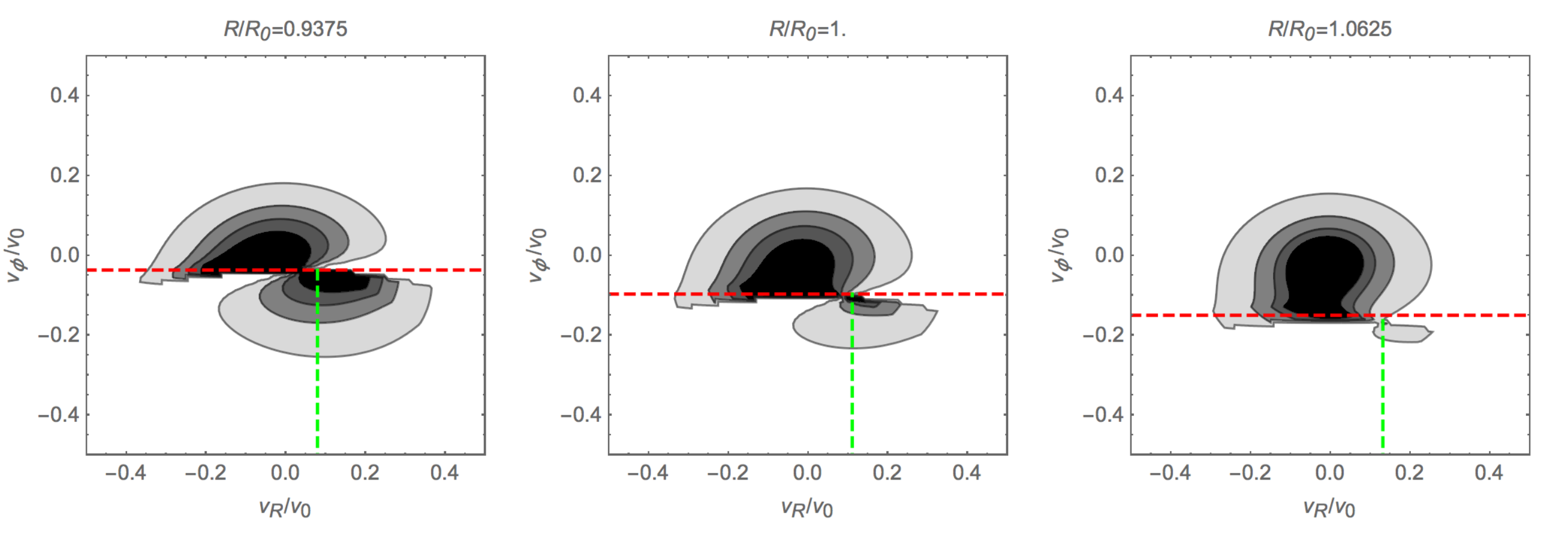}
  \caption{Theoretical shape of velocity space as a function of radius for a population of intrinsic radial velocity dispersion $\sim 35\kmsec$ in the presence of a bar with $\Omb=1.89\Omega_0$, angle from the long axis of the bar $\phi=25\degr$, and the maximum radial force being 1\% of the axisymmetric background, bf a potential corresponding to a flat rotation curve $\vc(R)=v_0$, computed using the formalism of \protect\cite{Monari2016,Monari2016c}, together with a red dashed horizontal line indicating $\vOLR$ computed as in \Eq{eq:vOLR}, and a green dashed vertical line representing the peak of the low-velocity mode with $v_\phi<\vOLR(R)$. The contours include (from the inner to the outer) 34, 50, 68, and 90 per cent of the stars.}
\label{fig:mod}
\end{figure*}

\begin{figure*}
  \centering 
 \includegraphics[width=0.99\textwidth]{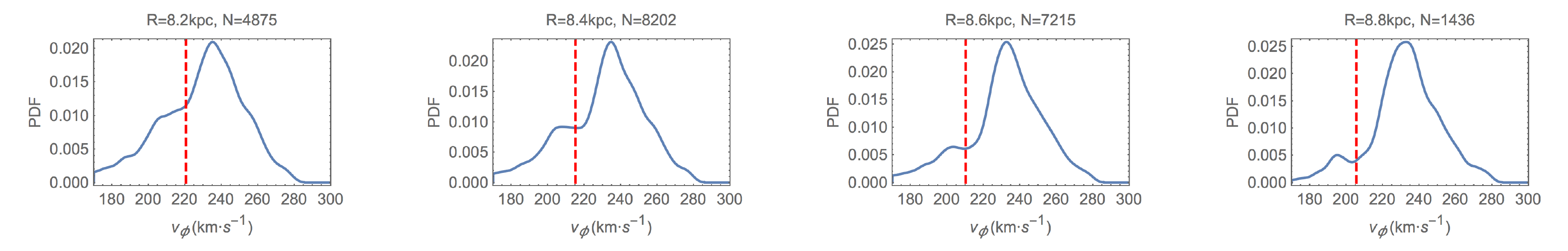}
  \caption{Distribution of stars in $v_\phi$, for stars with $v_R>0$
    and $R_i-\Delta R<R<R_i+\Delta R$, where $R_i=8.2\Kpc, 8.4\Kpc$,
    $8.6\Kpc$, and $8.8\Kpc$, and $\Delta R=0.1\Kpc$. The PDFs are
    obtained using Gaussian kernels of bandwidth $3\kmsec$. The red
    dashed line corresponds to $\vOLR(R_i)$ for
    $\Omb=1.89\Omega_0$, and $\beta=0$, indicating the theoretical gap
    between the high- and low-velocity modes in $v_\phi$.}
\label{fig:vphi}
\end{figure*}

\begin{figure*}
  \centering 
 \includegraphics[width=0.99\textwidth]{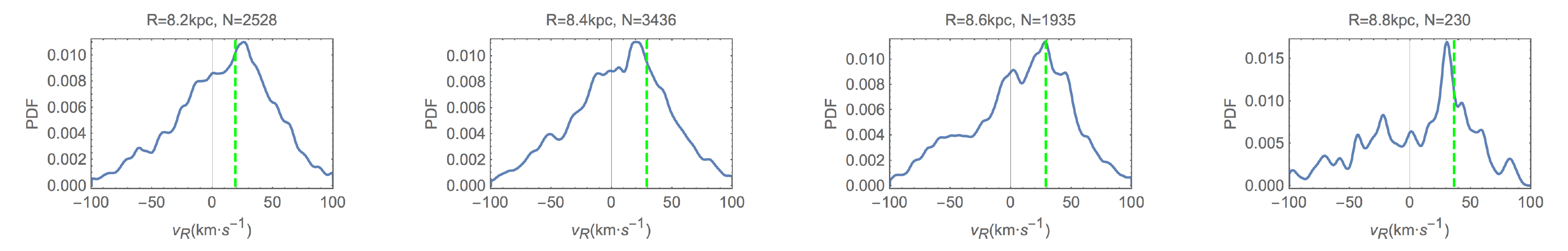}
  \caption{Distribution of stars in $v_R$, for stars with $v_\phi<\vOLR(R)$ and $R_i-\Delta R<R<R_i+\Delta R$, where $R_i=8.2\Kpc, 8.4\Kpc$, $8.6\Kpc$, and $8.8\Kpc$, and $\Delta R=0.1\Kpc$. The PDFs are obtained using Gaussian kernels of bandwidth $3\kmsec$. The green dashed lines corresponds to the peak of the low velocity mode at each $R_i$ from the theoretical model of \Fig{fig:mod}.}
\label{fig:vR}
\end{figure*}

To study the sample, we transform the observables $(\alpha,\delta,\pi,\mu_\alpha,\mu_\delta,\vlos)$ to the cylindrical Galactocentic coordinate system $(R,\phi,z,v_R,v_\phi,v_z)$. These transformations can be found in several textbooks \citep[e.g.][]{BinneyMerrifield}, and require us to know three fundamental parameters of our Galaxy: the Galactocentric radius of the Sun $R_0$, the circular speed of the Galaxy at the Sun $v_0$, and the motion of the Sun with respect to the Local Standard of Rest. The latter has a rather uncertain tangential component $V_\odot$ \citep[see][]{BT2008}\footnote{Each of the fundamental parameters of our Galaxy is still rather uncertain, but their combination $(v_0+V_\odot)/R_0$ (i.e. the angular speed of the Sun around the Galactic centre) is well constrained to $30.57\pm0.43\kmseckpc$ \citep{Reid2014}.}. We use the values found in \cite{Reid2014}, i.e. $R_0=8.34\Kpc$, $v_0=240\kmsec$, and $(v_0+V_\odot)=254.6\kmsec$. To determine $v_R$ and $v_\phi$, we also use $U_\odot=10\kmsec$ \citep[e.g.][]{Bovy2015}, the radial motion of the Sun with respect to the Local Standard of Rest\footnote{Note that the $U_\odot$ value is almost uninfluential on the distribution of stars in $v_\phi$ that we find in this work.}.

We focus in particular on the distribution of stars in the $(R,v_\phi)$ plane, shown in \Fig{fig:Rvphi_cut} (left panel). 
In this space, \emph{Hercules} is the clump of stars localized between $v_\phi\sim 190\kmsec$, and $v_\phi\sim 200\kmsec$, slightly detached from the main velocity mode at higher $v_\phi$. Notice the underdensity
of stars between \emph{Hercules} and the main velocity mode (the `gap'), which appears slightly inclined in the $(R,v_\phi)$ space. \emph{Hercules} and the gap disappear when we select only stars moving inwards with $v_R<0$ (\Fig{fig:Rvphi_cut}, central panel), while they are more evident when we select stars with $v_R>0$ (right panel). This is due to the fact that \emph{Hercules} is composed by stars moving outwards in the Galaxy \citep[e.g.][]{Dehnen1998,Famaey2005}.

If \emph{Hercules} is due to the effect of the bar's OLR, the gap's location in the $(R,v_\phi)$ plane corresponds to stars having their guiding radii\footnote{This is, strictly speaking, true only when a star's orbital eccentricity is null, but still a good approximation for the range of radial energies that we consider in our samples.} at $\ROLR$. At a radius $R$, the $v_\phi$ of these orbits is given by
\begin{equation}\label{eq:vOLR}
	\vOLR=\frac{\ROLR\vc(\ROLR)}{R}.
\end{equation}
This means that the angular momentum corresponding to the gap in velocity space at radius $R$ corresponds to the angular momentum of circular orbits at the OLR. As we move outwards in $R$, $\vOLR$ becomes lower, and the number of stars that is affected by the OLR becomes smaller, which eventually leads to the disappearance of the \emph{Hercules} moving group. We show this in \Fig{fig:mod} where, using the formalism of \cite{Monari2016,Monari2016c}, we compute the position of the \emph{Hercules} gap for perturbed phase-space distribution functions in the presence of a quadrupole bar perturbation. We show the theoretical shape of velocity space as a function of radius, for a population of intrinsic radial velocity dispersion $\sim 35\kmsec$, in the presence of a quadrupole bar\footnote{Higher-order terms with $m>2$ would in principle also appear in the expansion of the bar potential for more complex bar shapes than a pure quadrupole. The response to these additional terms in the potential will however be of second or higher order compared to the quadrupole amplitude, and will not affect the $\vOLR$ of the gap.} \citep[see Eq.~9 of][]{Monari2016c} with $\Omb=1.89\Omega_0$ (with $\Omega_0\equiv v_0/R_0$) and the maximum radial force being 1\% of the axisymmetric background, here an axisymmetric potential corresponding to a flat rotation curve $\vc(R)=v_0$. We also plot in \Fig{fig:mod} a red dashed horizontal line indicating $\vOLR$ computed as in \Eq{eq:vOLR}. In \Fig{fig:Rvphi_cut} (right panel)  \emph{Hercules} seems to disappear at $R\sim8.6\Kpc$. We can however still trace it at larger $R$ and lower $\vOLR$, using the $v_\phi$ distribution in \Fig{fig:vphi}. 

If we describe the circular velocity curve as a power-law $\vc(R)=v_0(R/R_0)^\beta$ in proximity of the Sun, then 
\begin{equation}
\ROLR(\Omb,\beta)=R_0\paresq{\frac{\Omega_0}{\Omb}\pare{1+\sqrt{\frac{1+\beta}{2}}}}^{1/(1-\beta)},
\end{equation}
\citep{Dehnen2000}. In the right panel of \Fig{fig:Rvphi_cut}, we plot the curve $\vOLR(R)$ for three values of $\ROLR$, using $\beta=-0.3,0,0.3$, and $\Omb=1.89\Omega_0$. The pattern speed $\Omb$ was found by \cite{Antoja2014} using the RAVE catalogue which mostly probes regions with $R<R_0$. We find that the three curves nicely follow the shape of the gap even for $R>R_0$, without {\it any} tuning of the parameters to obtain a good fit. This appears even more evident when looking at \Fig{fig:vphi}, where the $\vOLR$ for the flat rotation curve model (vertical red dashed line) perfectly matches, at every $R$, the saddle point in the $v_\phi$ distribution, corresponding to the \emph{Hercules} gap. To check the consequence of our choice of fundamental parameters for the Milky Way, we also studied the $(R,v_\phi)$ distribution for stars with $v_R>0$, but for the parameters of \citet{Bovy2015}, namely $R_0=8\Kpc$, $v_0=218\kmsec$, and $V_\odot=24\kmsec$. In this case, a slightly lower pattern speed of the bar relative to the local circular frequency is preferred ($\Omb\approx1.85\Omega_0$).

Ideally, one should not only compare the theoretical position of the gap in $v_\phi$ with the data, but also the detailed shape of velocity space. Indeed, the main conclusion from the present analysis is that the \emph{Hercules} moving group is generated by a perturbation with a single resonance radius at $R \sim 7$~kpc. This readily excludes corotating spiral arms with a varying pattern speed as a possible explanation. However, one should also compare the $v_R$ distribution of the analytical models to the data. It is not straightforward to make a direct quantitative comparison of the amplitude of the overdensity in the data and models, because while the models are reliably predicting the location of the overdensity in velocity space, they do not predict its amplitude due to the need for a refined treatment of the analysis at the resonance itself (Monari et al. in preparation). However, we can indeed compare the location of the overdensity in $v_R$ for stars in the low-velocity mode in the models and data. This is done in \Fig{fig:vR}, where the $v_R$ distribution as a function of radius is compared to the peak of the low velocity mode from the theoretical model of \Fig{fig:mod}: the agreement is good, given that the analytical model is precisely predicting the location of the gap in $v_\phi$ but is not a perfect representation of the actual distribution at the resonance.

\section{Conclusions}

The \emph{Hercules} moving group has traditionally been interpreted as a direct signature in local stellar kinematics of the Galactic bar's OLR \citep{Dehnen1999,Dehnen2000}. This explanation is however at odds with the slowly rotating long bar favoured by stellar and gas kinematics in the inner Galaxy. This would mean that an alternative explanation, e.g. based on spiral arms, should be found to reproduce the bimodality in local velocity space \citep{Monari2016c}. Such an alternative explanation has not yet been found, but \citet{Grand2014} showed that the outward radial migrators behind their corotating spiral arms display lower tangential velocities and an outward velocity, which could help explaining the moving group. However, a consequence of such a model would be that there is no unique resonant radius associated to the perturbation.

One way to test whether \emph{Hercules} is indeed linked to the OLR of the bar is to trace its position in velocity space as a function of position in the Galaxy. Here, we showed that this is already possible today through the combination of the TGAS and LAMOST DR2 catalogues. We found out that the \emph{Hercules} moving group is indeed closely following the prediction of models placing the Sun just outside the OLR of the bar. In these models, the corotation of the bar is close to $R \sim 4$~kpc, and its OLR is at $R \sim$7~kpc. This would mean that alternative explanations, necessary to account for a slowly rotating bar with corotation around $R\sim 6$~kpc, would not only have to reproduce the position of \emph{Hercules} in local velocity space at the Sun's position, but also its variation with radius precisely as predicted by fast bar models. As the observed variation with radius of the gap in $v_{\phi}$ indicates, the \emph{Hercules} moving group is linked to a single resonance radius, hence excluding a varying pattern speed with radius. This excludes a corotating spiral origin for the group. What is more, its observed variation in $v_R$ as a function of radius also closely follows the prediction of the bar model, rendering an explanation based on a spiral density wave unlikely.

The deprojected 3D density of red clump giants in the inner Galaxy nevertheless
points to the existence of a long, flat structure, in the direct prolongation of the bar, reaching out to $R \sim 5$~kpc. If the bar indeed has a high pattern speed $>1.8 \Omega_0$, then this long thin structure cannot be rotating with the bar, but could rather be a loosely wound spiral coupled to the end of the bar. It will thus be of extreme importance to model precisely the effect of spiral
arms in the inner Galaxy, and to explore their whole possible parameter space, to check whether spirals could help reconcile existing data on gas and stellar kinematics in the inner Galaxy with the fast bar revealed by the stellar kinematics in the outer Galaxy. A second possibility would be that the long extension of the bar and the smaller bar causing the OLR feature in the Solar vicinity actually constitute a double-bar feature, the smaller bar having a larger pattern speed \citep[e.g.][]{Wozniak1995}. A last intriguing possibility would be that the pattern speed of the bar has dramatically decreased in the recent past, and that the outer Galactic disc is still affected by its ancient pattern speed.

\section*{Acknowledgements}
We thank James Binney for useful discussions. Most of this work has been realised during the IVth Gaia Challenge workshop (Stockholm, 2016). This work has made use of data from the European Space Agency (ESA) mission {\it Gaia} (\url{http://www.cosmos.esa.int/gaia}), processed by the {\it Gaia} Data Processing and Analysis Consortium (DPAC, \url{http://www.cosmos.esa.int/web/gaia/dpac/consortium}). Funding
for the DPAC has been provided by national institutions, in particular the institutions participating in the {\it Gaia} Multilateral Agreement.
Guoshoujing Telescope (the Large Sky Area Multi-Object Fiber Spectroscopic Telescope LAMOST) is a National Major Scientific Project built by the Chinese Academy of Sciences. Funding for the project has been provided by the National Development and Reform Commission. LAMOST is operated and managed by the National Astronomical Observatories, Chinese Academy of Sciences.




\bibliographystyle{mn2e}
\bibliography{mn-jour,lamtgasbib}
\label{lastpage}
\end{document}